\documentstyle[prl,aps,multicol]{revtex}

\begin{document}
\draft
\title{Pseudogap in the Optical Spectra of  UPd$_2$Al$_3$}
\author{M. Dressel$^{1,}$\cite{email}, B. Gorshunov$^{1,}$\cite{permanent1}, 
N. Kasper$^{1,}$\cite{permanent2}, B. Nebendahl$^{1}$, 
M. Huth$^{2}$, and H. Adrian$^{2}$}
\address{$^{1}$ Physikalisches Institut, Universit\"at Stuttgart,
Pfaffenwaldring 57, D-70550 Stuttgart, Germany\\
$^2$ Institut f\"ur Physik, Universit\"at Mainz, 
Staudinger Weg 7, D-55099 Mainz, Germany}
\date{Received \today}
\maketitle

\begin{abstract} 
The in-plane optical conductivity of UPd$_2$Al$_3$ was measured at 
temperatures $2~{\rm K}<T<300$~K in the spectral range from 
1~cm$^{-1}$ to 40~cm$^{-1}$ (0.14~meV to 5~meV). As the temperature 
decreases below 25~K a well pronounced pseudogap of $0.2$~meV 
develops in the optical response. In addition we observe a narrow 
conductivity peak at zero frequency which at 2~K is less than 
1~cm$^{-1}$ wide but which contains only a fraction of the 
delocalized carriers. The gap in the electronic excitations might be 
an inherent feature of the heavy fermioin ground state. 
\end{abstract}

\pacs{PACS numbers: 71.27.+a, 74.70.Tx, 78.20.--e}

\begin{multicols}{2}
\columnseprule 0pt
\narrowtext


The heavy fermion (HF) compound UPd$_2$Al$_3$ is one of the few 
materials which shows coexistence of superconductivity and magnetic 
ordering \cite{Geibel91}. Susceptibility as well as resistivity data 
indicate the formation of a coherent state below a characteristic 
Kondo-lattice temperature $T^{*}\approx 20$~K; commensurate 
antiferromagnetic (AF) order develops below $T_{\rm N}=14$~K, 
superconductivity sets in at $T_c=2$~K. For $2~{\rm K}<T<14$~K the 
electronic specific heat shows a $C_{el}/T\propto T^2$ dependence; 
the effective mass of  the charge carriers is estimated as 
$m^*/m_0\approx 50$ \cite{Geibel91,Dalichaouch92} indicating 
strongly correlated electronic states in coexistence with the 
long-range AF order. It was suggested \cite{Caspary93} that two 
different groups of uranium $5f$ states exist in UPd$_{2}$Al$_3$: one 
group is assumed to be rather localized and responsible for the 
magnetic properties whereas the states of another sort are less 
localized and responsible for the superconductivity of the compound.

Optical experiments have proven to be sensitive to the formation of 
heavy quasiparticles \cite{Degiorgi99}. In the HF compounds above 
$T^*$ the optical conductivity reveals a broad Drude behavior which 
characterizes normal metals. Below $T^{*}$ the increase in the 
density of states (DOS) at low energies is commonly described  by 
the formation of  a narrow Drude-like response with a renormalized 
scattering rate $\Gamma^*$ \cite{Fukuyama85}. The signature of 
magnetic ordering in the optical response of heavy fermions was 
studied in Ref.~\onlinecite{Degiorgi94}. In spin-density-wave (SDW) 
systems like URu$_2$Si$_2$, the opening of a single-particle 
gap in the electronic DOS at the Fermi surface is clearly seen in the 
optical properties \cite{Degiorgi94,Bonn88}. No 
effect of the magnetic ordering was found in the infrared 
response of  UPd$_2$Al$_3$ so far. Neutron scattering 
results \cite{Krimmel92} reveal rather large ordered 
moments which reside predominantly at the uranium sites. Based on this 
experimental evidence, UPd$_2$Al$_3$ was 
described as a local-moment magnet and hence, the magnetic 
ordering should have only a minor influence on the electronic DOS. 
In this Letter we report on optical experiments at
millimeter-submillimeter (mm-submm) wavelengths which clearly show 
the development of a pseudogap in the DOS in UPd$_2$Al$_3$ below 25~K; 
our preliminary study on 
another film \cite{Dressel97} showed the same features in the optical response;
these findings cannot be easily explained by  existing models.


The highly $c$-axis oriented thin (150~nm) film of UPd$_2$Al$_3$ was 
prepared on (111) oriented LaAlO$_3$ substrate (thickness 0.924~mm) 
by electron-beam co-evaporation of the constituent elements in a 
molecular-beam epitaxy system \cite{Huth93}. The phase purity and 
structure of the film were investigated by X-ray and reflection 
high-energy electron diffraction. The high quality of the film is 
seen in dc resistivity data displayed in Fig.\ \ref{fig:figure1} 
which  perfectly agree with measurements of single crystals 
\cite{Geibel91}. 

For the measurements a coherent source mm-submm spectrometer was 
employed \cite{Volkov85} utilizing a set of backward wave 
oscillators as monochromatic but continuously tunable sources. The 
use of a Mach-Zehnder arrangement allows for measuring both the 
amplitude and the phase of the signal transmitted through the 
sample, which in our case is a film on a substrate (with the electric 
field $\bf E$ perpendicular to the $c$-axis). From these two 
quantities the conductivity $\sigma(\omega)$ and dielectric constant 
$\epsilon(\omega)$ of the film are evaluated using Fresnel's 
formulas for a two-layer system \cite{Born80} without assuming any 
particular model. The optical parameters of the substrate are 
determined beforehand by performing the experiments on a blank 
substrate. The large size of the sample ($1\times 1~{\rm cm}^{2}$) 
allowed us to extend the measurements to very low frequencies, from 
$\omega/(2\pi c)= 40$~cm$^{-1}$ down to 1.15~cm$^{-1}$. In addition, 
we performed microwave experiments at 10~GHz using a setup designed 
for measuring superconducting films \cite{Peligrad98}. A small slice 
of a sample ($4\times 1~{\rm mm}^2$) was placed in the electric 
field maximum of a cylindrical cavity (${\bf E}\perp c$); we obtain 
the conductivity (Fig.~\ref{fig:figure1}) and dielectric constant 
from the change in width and shift in frequency using cavity 
perturbation theory.

In Fig.\ \ref{fig:figure2} we present the raw transmission spectra 
for the UPd$_2$Al$_3$ film on the substrate for two selected 
temperatures (100~K and 2~K) which illustrate clearly our main 
finding. The fringes in the spectra are due to multi-reflection of 
the radiation within the plane-parallel substrate acting as a 
Fabry-Perot resonator for the monochromatic radiation \cite{Born80}. 
The distance between the peaks is mainly determined by the thickness 
and refractive index of the substrate, their amplitude by the 
parameters of the film. At  $T\geq30$~K the transmission maxima and 
minima only slightly depend upon  frequency, implying a frequency 
independent conductivity and dielectric constant of the 
UPd$_2$Al$_3$ film. In contrast, the overall transmission for 
lower temperatures is strongly reduced below $10~{\rm cm}^{-1}$ due to  
absorption within the film. At lower frequencies, below 3.5~${\rm 
cm}^{-1}$, the transmitted signal increases again, indicating an 
absorption edge. This feature can also be seen by looking at the 
temperature dependencies of the transmission and the conductivity as 
plotted in Fig.\ \ref{fig:figure1} for several fixed frequencies.

The conductivity and dielectric constant of UPd$_2$Al$_3$ are 
plotted in Fig.~\ref{fig:figure3} as functions of frequency for 
different temperatures. Within our accuracy they are frequency 
independent for $T\geq30$~K which resembles a low-frequency response 
of a Drude metal; the value of the conductivity agrees well with the 
dc result. It is clearly seen that for low temperatures the optical 
response of UPd$_2$Al$_3$ differs from that of a renormalized Drude 
metal. The simple Hagen-Rubens relation in general used to 
extrapolate the far-infrared reflectivity below 30~cm$^{-1}$ 
\cite{Degiorgi99} totally misses the two features in the mm-submm 
range. First, the optical conductivity $\sigma(\omega)$ clearly 
shows the development of a minimum below 3~cm$^{-1}$~at $T<30$~K 
with a correspondent pronounced increase of $\epsilon(\omega)$. We 
ascribe this minimum to the opening of a pseudogap in the electronic 
DOS (see below). This feature gradually disappears with increasing 
temperature and is not seen above 30~K. Second, at frequencies below 
approximately 1.5~cm$^{-1}$ the conductivity increases towards 
considerably higher dc values leading to a very narrow peak at 
$\omega=0$. With the help of our microwave data, we estimate its 
width to be 0.3~cm$^{-1}$ at $T=2$~K.


To discuss our findings let us turn back to 
Fig.\ \ref{fig:figure1}. Below $T_{\rm N}$ the temperature dependence 
of the dc resistivity of the film can be described using the 
expression for an antiferromagnet with energy gap $E_g$ 
\cite{Dalichaouch92,Bakker93}:  
\begin{equation}
\rho(T)=\rho_0 + aT^2 + bT\left(1+\frac{2k_{\rm B}T}{E_g}\right)
\exp\left\{-\frac{E_g}{k_{\rm B}T}\right\}.
\label{eq:magnon}
\end{equation}
Here $\rho_0$ gives the residual resistivity and the second term 
indicates the electron-electron scattering. Fitting the dc data of 
Fig.\ \ref{fig:figure2} yields a gap value $E_g=1.9$~meV, which 
corresponds to the one reported in
\cite{Dalichaouch92,Huth93,Bakker93}. From point contact 
measurements a larger gap  up to 12.4~meV was 
suggested \cite{Aarts94}. Using thin films of UPd$_2$Al$_3$ recent 
tunneling experiments in the superconducting and normal state 
\cite{Jourdan99} clearly demonstrate that a reduction of the DOS 
remains in the temperature range roughly up to $T_{\rm N}$. The existence 
of a gap in either the electronic DOS or in the magnon spectrum was 
opposed by Caspary {\it et al.} \cite{Caspary93} and on the ground of 
infrared measurements by Degiorgi {\it et al.} \cite{Degiorgi94}. In 
the light of the low-energy measurements presented in this Letter 
the arguments will be reconsidered below.

To analyze the spectral weight of the optical response, we first 
neglect the pseudogap and describe the 
mm-submm spectra of $\sigma(\omega)$ and $\epsilon(\omega)$ in terms 
of a renormalized Drude response (dotted line in Fig.~\ref{fig:figure4}) 
\begin{equation}
\sigma(\omega)= 
\frac{\Gamma^*}{4\pi}\frac{(\omega_p^{*})^2}{\omega^2+(\Gamma^{*})^2} 
\quad , 
\quad \epsilon(\omega)= 1-
\frac{(\omega_p^{*})^2}{\omega^2+(\Gamma^{*})^2} 
\label{eq:Drude}
\end{equation}
where the plasma frequency is related to the carrier density $n$ and 
the effective mass $m^*$ by $\omega_p^*=\sqrt{4\pi ne^2/m^*}$. At 
$T=2$~K we find $\omega_p^*/(2\pi c)\approx 4300~{\rm 
cm}^{-1}$. Assuming that the total number of charge carriers 
remains unchanged, sum-rule arguments give $\omega_p/\omega^*_p = 
\sqrt{m^*/m_b}$ where $\omega_p$ is the unrenormalized plasma 
frequency and $m_b$ the band mass. With $\omega_p/(2\pi c)= 
44\,000~{\rm cm}^{-1}$  \cite{Degiorgi94} 
we obtain $m^*/m_b\approx 105$.  Using experimental \cite{Inada1} 
and calculated \cite{Sandratskii94} de Haas-van Alphen spectra we 
know that approximately 35~\%\ of the electrons contribute to the HF 
state, leading to the effective mass of $m^*/m_b=40$. 

In contrast to the above description by a renormalized Drude metal, 
we find clear evidence for a narrow pseudogap which develops in 
UPd$_2$Al$_3$ at low temperatures  reducing the spectral weight in 
the conductivity spectrum and correspondingly the plasma frequency 
$\omega_p^*$. At 2~K this reduction amounts in 40\%\ (hatched area 
in Fig.4) leading 
finally to the effective mass $m^*/m_b=60$ which agrees well with 
the value obtained by thermodynamic methods ($m^*/m_0=41-66$) 
\cite{Geibel91,Dalichaouch92}, but is a somewhat  smaller than those 
got from previous infrared measurements ($m^*/m_b=85$) 
\cite{Degiorgi94}. The gap value can be determined using a  
semiconductor approach with $\sigma(\omega) \propto \omega^{-1} 
\sqrt{\hbar\omega- E_g}$ giving $E_g\approx0.23$~meV (solid line in 
Fig.\ \ref{fig:figure4}). This value is essentially temperature 
independent (Fig.\ \ref{fig:figure3}a) in contrast to the  
BCS-like temperature behavior of the energy gap in superconducting 
or density-wave ground systems. Instead it becomes more and more 
pronounced as the temperature is lowered. 

The most interesting question is the relation between the AF 
ordering and the pseudogap. The fact that we also see the gap-like 
feature above $T_{\rm N}$ up to 30~K does not rule out its 
connection to the magnetic ordering since an incommensurate phase 
was observed in UPd$_2$Al$_3$ up to $T\approx20$~K by neutron 
diffraction experiments \cite{Krimmel92}, and a maximum of the 
magnetic susceptibility appears  around 35~K \cite{Geibel91}. We 
want to discuss three possible explanations: (i) the pseudogap in 
the optical response may be related to spin-wave excitations; (ii) 
the formation of a SDW ground state may lead to the opening of an 
energy gap. (iii) On the other hand since $T^*$ is also in this 
temperature range, the gap in the electronic DOS might be a 
property of the coherent HF ground state. 

(i) Inelastic neutron scattering experiments found two 
very-low-energy modes, one at 1.5~meV which is ascribed to spin-wave 
excitations, and another at 0.4~meV which is associated with 
superconductivity \cite{Bernhoeft98}. These modes exhibit a strong 
$T$ and {\bf q}-dependence. Besides the resistivity $\rho(T)$ 
discussed above, the drop of $C_{el}(T)$ \cite{Geibel91,Caspary93} 
and torque measurement results of the magnetization \cite{Sullow96} 
were also interpreted as being due to the opening of a gap in the 
magnetic excitation spectrum. However, we exclude the possibility 
that the features discovered in our low-temperature mm-submm spectra 
are due to purely magnetic absorption since we were not able to 
obtain a reasonable fit of these features by a ``magnetic 
Lorentzian'' absorption line. We definitely see the development of a 
gap in the {\em electronic} excitation spectra. 

(ii) The low-temperature decrease of the resistivity of 
UPd$_2$Al$_3$ can be well described by an exponential behavior 
[Eq.~(\ref{eq:magnon})]. The  so-obtained  gap $E_g$=1.9 meV may 
indicate the formation of a SDW state like in URu$_2$Si$_2$ 
\cite{Dalichaouch92,Bakker93}. However, the gap energy 0.23~meV 
obtained from our optical results is about an order of magnitude 
smaller; it is also a factor of 20 below the value one would expect 
from mean field theory $E_g = 3.5k_{\rm B}T_{\rm N}\approx 4.2$~meV. 
Also the comparison of NMR and NQR results for UPd$_2$Al$_3$ 
\cite{Tou95} and for the SDW systems URu$_2$Si$_2$ \cite{Kohara86} 
and UNi$_2$Al$_2$ \cite{note} does not corroborate the SDW picture 
for UPd$_2$Al$_3$. The itinerant antiferromagnetism of a SDW also 
is at odds  with the formation of local moment magnetism 
deduced from susceptibility \cite{Geibel91} and neutron scattering 
\cite{Krimmel92}. 

(iii) It was suggested \cite{Caspary93,Sandratskii94} that two 
electronic subsystems coexist in UPd$_2$Al$_3$. One of them is a 
rather localized uranium $5f$ state responsible for the magnetic 
properties, the other part is delocalized and determines the HF and 
superconducting properties. From the London penetration depth 
$\lambda_L(0)=450$~nm \cite{Caspary93} we calculate
\cite{Tinkham96} the plasma 
frequency of superconducting carriers $(2\pi\lambda_L)^{-1} = 
3540~{\rm cm}^{-1}$ and find a perfect agreement with 
$\omega_p^*/(2\pi c) = 3500~{\rm cm}^{-1}$ obtained from the 
spectral weight under our mm-submm conductivity spectrum just above 
$T_c$ (grey area in Fig.\ \ref{fig:figure4}). This 
not only is an independent confirmation of the pseudogap but it also 
implies that all carriers seen in our spectra are in the HF ground 
state and eventually undergo the superconducting transition below 
$T_c$. We can definitely rule out an assignment of the gap to 
the localized carriers of the AF ordered states 
with the delocalized carriers contributing only to the narrow feature 
at $\omega=0$ because with a plasma frequency of 1500~cm$^{-1}$ at 
low temperatures, it accounts for only 18\%\ of the carriers which 
become superconducting. This means that also the excitations above 
the gap stem from the delocalized states and that the pseudogap 
observed in our conductivity spectra is either inherent to the heavy 
quasiparticle state or it is related to magnetic correlations of the 
second subsystem.


In conclusion, the electrodynamic response of UPd$_2$Al$_3$ in the 
low energy range from 0.14 meV to 5 meV (1~cm$^{-1}$ to 40 
cm$^{-1}$) exhibits a behavior at low temperatures ($T\leq 25~K$) 
which cannot be explained within the simple picture of a 
renormalized Fermi liquid. We observe an extremely narrow (less than 
0.1~meV) Drude-like peak at $\omega=0$ (dashed line 
in Fig.~\ref{fig:figure4}) and 
a pseudogap of about 0.2~meV. The experiments yield indications that 
this pseudogap is not a simple SDW gap but rather is connected to 
correlations of the delocalized carriers, and  may be a general 
signature of HF compounds. Recent millimeter wave experiments on 
UPt$_3$ \cite{Donovan97} which show a peak in the conductivity at 
6~cm$^{-1}$ for temperatures below 5~K indicate a similar scenario.  
While the finite-energy excitations occur at comparable frequency in 
UPd$_2$Al$_3$ and in UPt$_3$, the energy scale of the magnetic 
ordering is somewhat different, since $T_{\rm N}\approx 14$~K in 
UPd$_2$Al$_3$ while magnetic correlations are found in UPt$_3$ only 
below 5~K \cite{Kjems88}. Investigations of the magnetic field 
dependence which might shine light on this problem are in progress.

We thank G. Gr{\"u}ner, A. Loidl, M. Mehring and A.Mukhin for helpful 
discussions. The work was partially supported by the 
Deutsche Forschungsgemeinschaft (DFG) via Dr228/9 and SFB~252.

\begin{figure}
\caption{\label{fig:figure1}
Temperature dependence of the dc and ac conductivities of a
UPd$_2$Al$_3$ film. At $T_{\rm N}$ the systems orders
antiferromagnetically, at $T_c$ it becomes superconducting.
The solid line represents a fit using 
Eq.~(\protect\ref{eq:magnon}) with the gap $E_g=1.9$~meV. The inset 
shows the temperature dependence of the transmission through film 
on LaAlO$_3$ substrate for different frequencies. 
Dashed lines are guides to the eye.}   
\end{figure}

\begin{figure}
\caption{\label{fig:figure2}
Spectra of transmission of a 150~nm thick UPd$_2$Al$_3$ film on 
LaAlO$_3$ (thickness 0.924~mm) for two temperatures $T=100$~K  
and 2~K. The periodical maxima are caused by interference of the 
radiation in the substrate. The dashed lines  connecting 
minima and maxima are drawn to emphasize the overall frequency 
dependence of the transmission of the film.}        
\end{figure}

\begin{figure}
\caption{\label{fig:figure3}
(a) The conductivity  and (b) the dielectric constant of 
UPd$_2$Al$_3$ as functions of frequency at several temperatures. 
The dc conductivities are shown on the left side of the upper panel;
the data at 0.33~cm$^{-1}$ are obtained by microwave cavity 
measurements. 
The inset shows the optical conductivity over a wide frequency range 
using the data of \protect\cite{Degiorgi94}. The lines are drawn to 
guide the eye.} 
\end{figure}

\begin{figure}
\caption{\label{fig:figure4}
Frequency dependent conductivity $\sigma(\omega)$ (dots) of 
UPd$_2$Al$_3$ at $T=2$~K. The dotted line shows a fit 
of the data with the 
renormalized  Drude conductivity [Eq.~(\protect\ref{eq:Drude})]. 
The dashed curve is a Drude description of the  
$\omega=0$ peak with a width of 0.3~cm$^{-1}$. 
The solid line indicates a 
$\protect\omega^{-1}\protect\sqrt{\hbar\omega - E_g }$ behavior of 
the pseudogap used for the absorption edge in semiconductors. }
\end{figure}

\end{multicols}
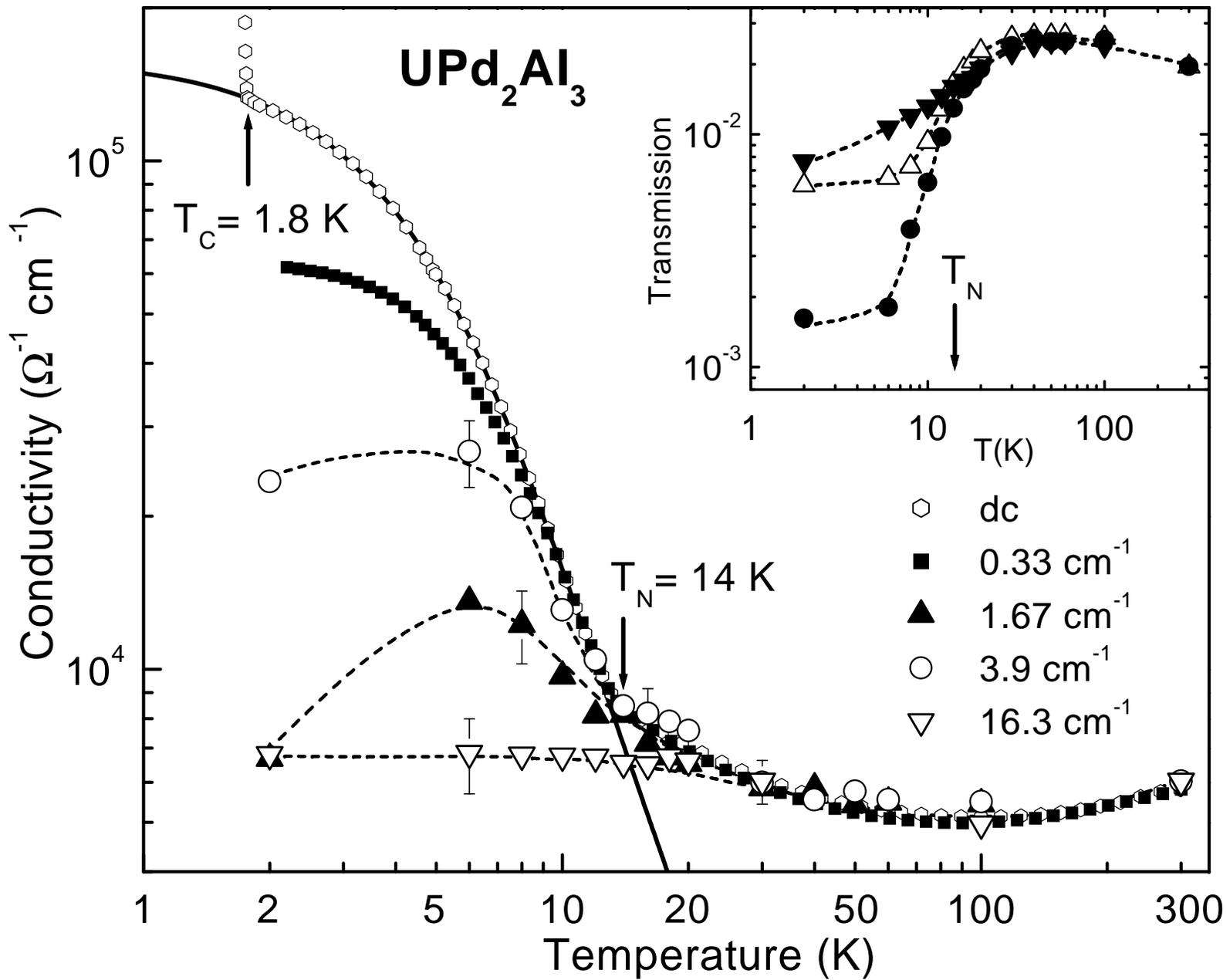
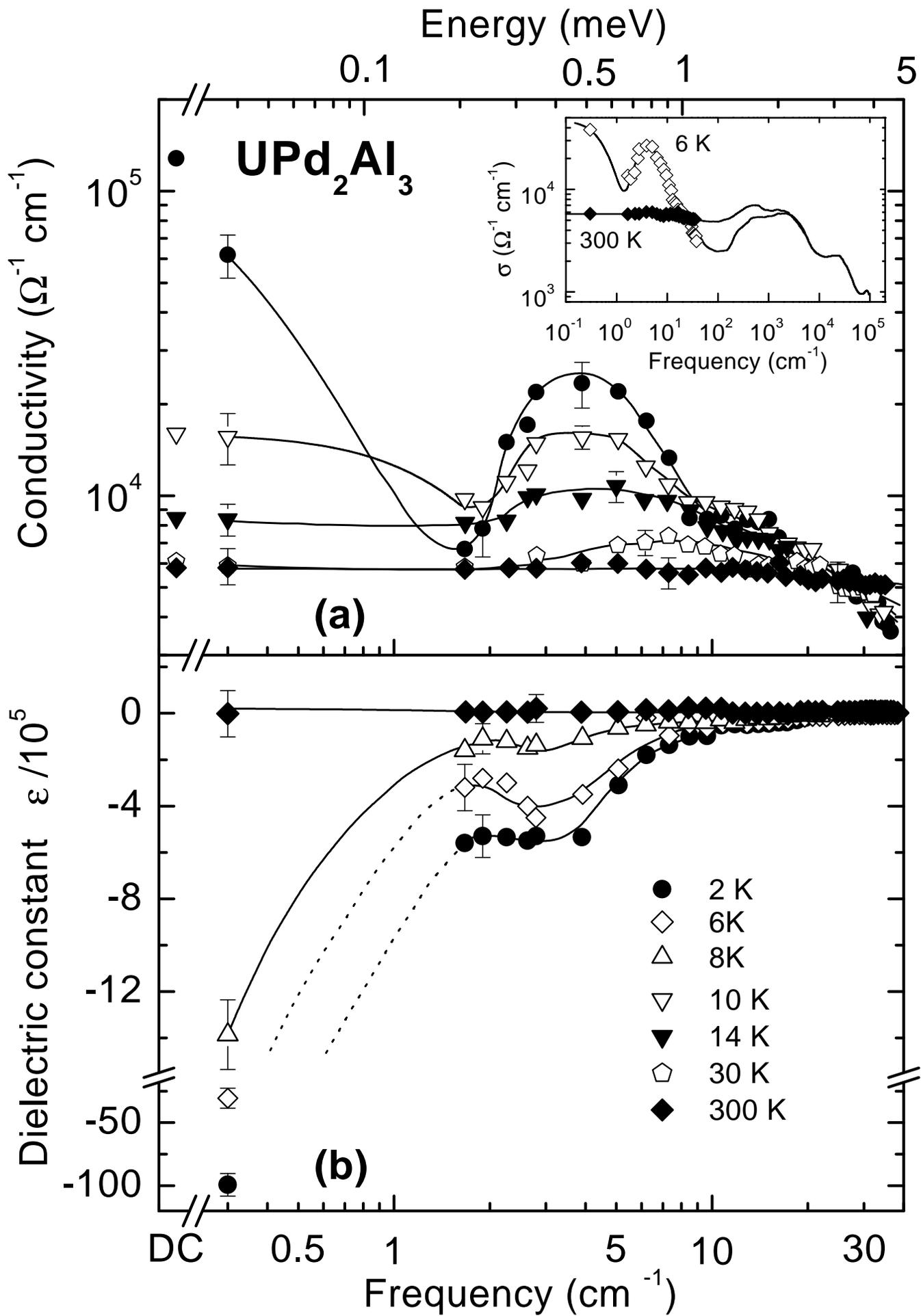
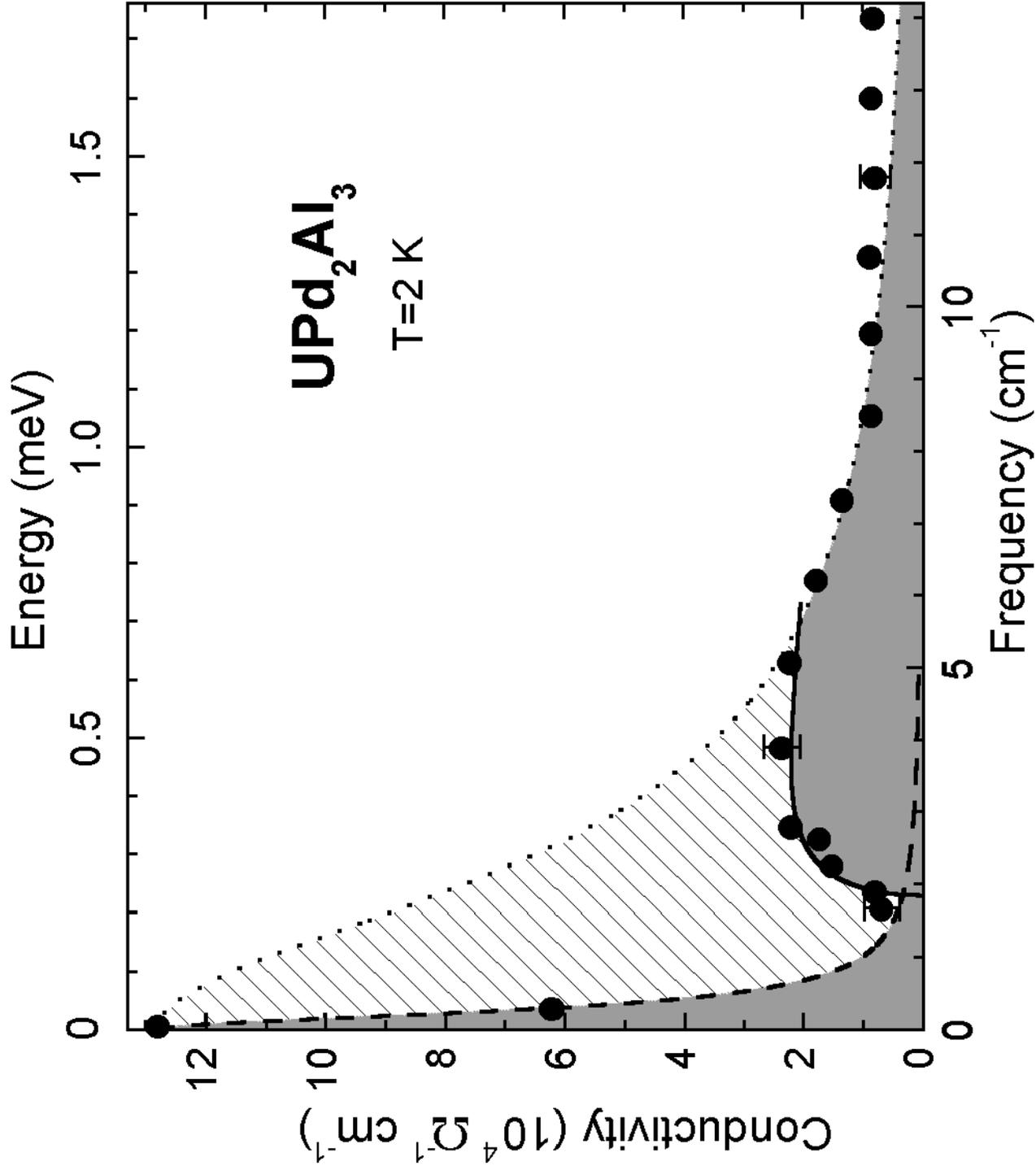
\end{document}